\documentclass[prl,twocolumn]{revtex4}

\usepackage{latexsym}
\usepackage{graphicx}

\begin{document}
\draft
\title{How Fast Elements can Affect Slow Dynamics} 
\author{Koichi Fujimoto\thanks{fujimoto@complex.c.u-tokyo.ac.jp}
 and Kunihiko Kaneko}
\address{Department of Pure and Applied Sciences, University of Tokyo,
Komaba, Meguro, Tokyo 153-8902, JAPAN}

\date{\today}

\begin{abstract}
A chain of coupled chaotic 
elements with different time scales is studied.  
In contrast with the adiabatic approximation, we find correlations
between faster and slower elements 
when the differences in the time scales of the elements 
lie within a certain range.  
For such correlations to occur, three features are essential:
strong correlations among the elements allowing for
both synchronization and desynchronization,
bifurcation in the dynamics of the fastest element by the change of its
control parameter, and  the cascade propagation of the bifurcation. 
The relevance of our results to biological memory is briefly discussed.
\end{abstract}

\maketitle

Many biological, geophysical 
and physical problems include a variety of modes with different time scales.
The study of dynamical systems with various time scales is 
important for understanding 
the hierarchical organization of such systems by investigating the dynamic interactions 
among modes. 
Adiabatic elimination\cite{Haken}
is often adopted for systems with different time scales.
If the correlations between modes with different scales are neglected,
the fast variables are eliminated and the dynamics of the system is
expressed only by the slow variables.  
The fast variables are then replaced by their averages and noise.
In this adiabatic approximation, the characteristics of the dynamics of the fast time
scales disappear, and 
the "information flow" from fast to slow time scales
is only retained as memory terms in a Langevin equation. 

The adiabatic approximation is valid when the
differences between the time scales are large.
However, when the differences are small,
correlations between the modes appear, invalidating the approximation.
Then,  the fast scale dynamics 
can influence the dynamics of slower variables.
Here we investigate under what conditions the faster variables
can influence the dynamics of the slower variables.
We will show that a dynamical
"information flow" from fast to slow dynamics is possible 
when a given condition is satisfied.  For this, 
chaos is relevant since it makes possible the
amplification of microscopic perturbations to a macroscopic scale. 
However, in order for the propagation of statistical properties from fast to slow
variables to occur, it turns out that two other properties are required: coherence and
a cascade of bifurcations.

In the present paper, we investigate how the
statistical (topological) properties of the slow dynamics can
depend on those of the fast dynamics
by adopting a coupled dynamical system with different time scales.
To be specific, 
we choose a chain of  nonlinear oscillators whose typical time scales
are distributed as a power series.
The dynamics of each oscillator is assumed to differ only in its time scale, and thus
there are only three  control parameters in our model:
one for the nonlinearity, one for the coupling strength among oscillators,
and one for the difference in time scales.

The concrete form adopted here is as follows:
We choose the Lorenz equation as the single oscillator,
\begin{equation}
\left\{\begin{array}{ll}
\dot{x} = f_x(\vec{X}) \equiv 10 ( y - x )\\
\dot{y} = f_y(\vec{X}) \equiv - x z + r x - y \\
\dot{z} = f_z(\vec{X}) \equiv x y -\frac{8}{3} z \
\end{array}\right.
\label{eq:Lorenz}
\end{equation}
where $\vec{X} \equiv (x, y, z)$.
The time scale differences are introduced as
\begin{equation}
T_i \frac{d \vec{X}_i}{d t} = \vec{F}(\vec{X}_i) , \ \ \ \ \
T_i \equiv T_1\tau^{i-1}
\label{timescale-diff}
\end{equation}
where
$\vec{F}(\vec{X}) = (f_x(\vec{X}), f_y(\vec{X}),f_z(\vec{X}))$.
The index of the elements is denoted as $i$ with $i = 1,2,,,L =$System size.
$T_i$ is the characteristic time scale for each element
and $\tau$ $( < 1 )$ is the time scale difference.
Using a power series distribution for the characteristic time scales
is analogous to the shell model for turbulence\cite{shell}.
The total time scale difference is given by
\begin{equation}
T_{total} \equiv \frac{T_L}{T_1} =  \tau^{L-1}.
\label{eq:totalgap}
\end{equation}
In the present Letter, we adopt the system size $L$
as a control parameter by fixing $T_{total} = 100$ and couple neighboring elements
diffusively as follows:
\begin{equation}
T_i \frac{d \vec{X}_i}{d t}  = \vec{F}( ({\bf E}-{\bf D}) \vec{X}_i + \frac{1}{2}
{\bf D} (\vec{X}_{i-1}+ \vec{X}_{i+1}) )
\label{couple}
\end{equation}
\[
{\bf D} = \left[
\begin{array}{ccc}
d_x & 0 & 0 \\
0 & d_y & 0 \\
0 & 0 & d_z
\end{array}
\right],
{\bf E} = \left[
\begin{array}{ccc}
1 & 0 & 0 \\
0 & 1 & 0 \\
0 & 0 & 1
\end{array}
\right]
\]
where we chose  $d_y = 0$, $d_x = d_z = d = 0.49$.
The Runge-Kutta method was used with a time step size
such that the fastest element $\vec{X}_i$ is computed with high precision.

\begin{figure}[ht]
\begin{center}
\includegraphics[width=8.6cm,height=3.2cm]{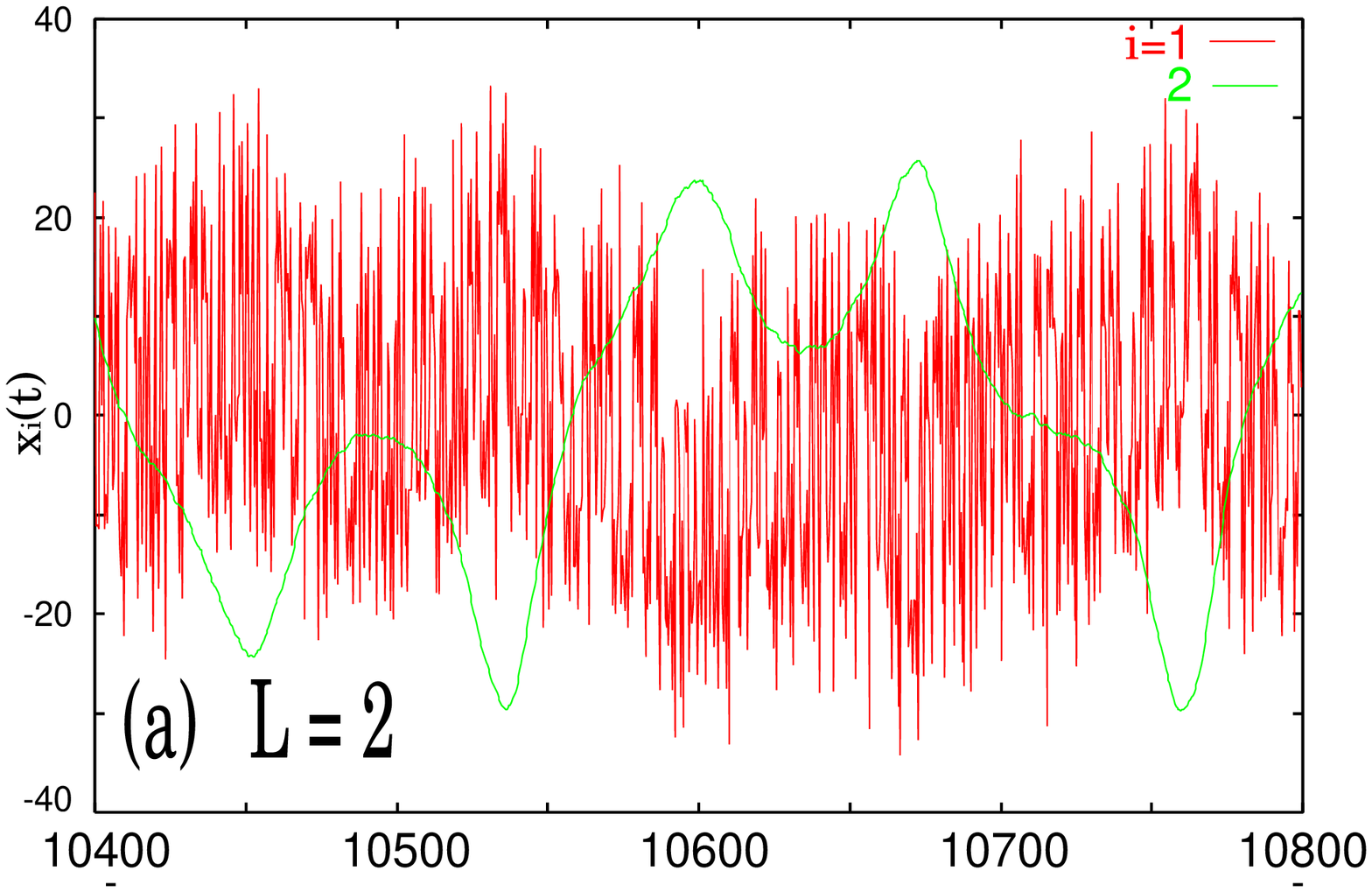} 
\includegraphics[width=8.6cm,height=3.2cm]{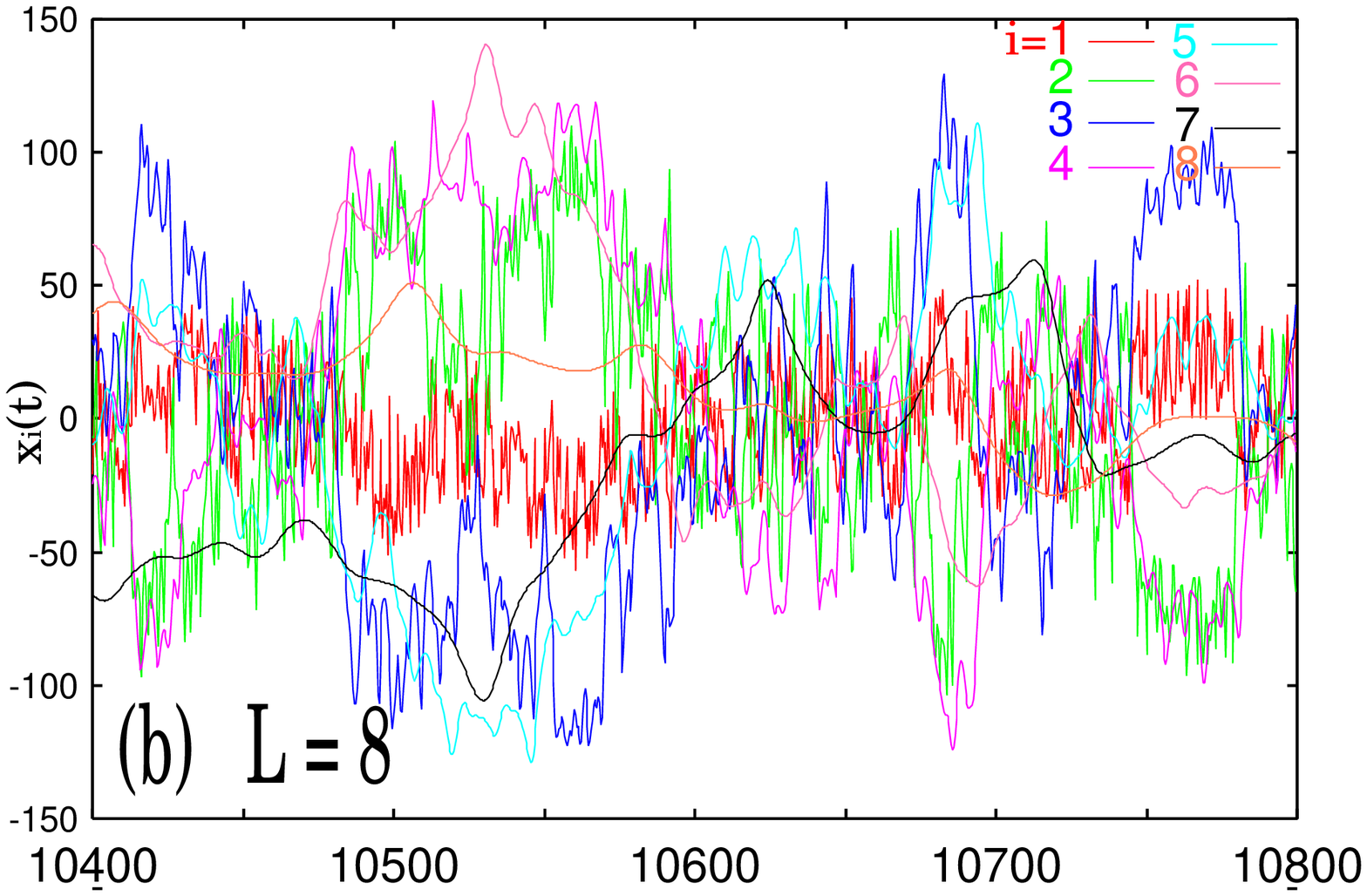} 
\caption{Time series of $x_i(t)$.
$(L, \tau)  =$ $(2, 100)$ (a) and  $(8, 1.93)$ (b).
(b) shows correlated motion of elements with various time scales.
The colors correspond to the element index.
$r = 41$ at $1 \leq i \leq L$ and $d = 0.49$.}
\label{fig:timeseries}
\end{center}
\end{figure}

Representative examples for the time series of $x_i(t)$ are plotted
in Fig.\ref{fig:timeseries}, with
$(L, \tau)  =$ $(2, 100)$ in (a)
and $(L, \tau) =$ $(8, 1.93)$ in (b).
In (a), there is no explicit correlation between
the dynamics of the fast time scale at $i = 1$ and the slow one at  $i = L = 2$.
On the other hand, there is phase synchronization\cite{Pikovsky} at various time scales in (b).
Here the phase relation between $x_i$ and $x_{i+1}$ is anti-phase.
The co-variance, given by
$\frac{\langle ( x_i - \langle x_i \rangle)( x_{i+1} - \langle
x_{i+1}\rangle ) \rangle}{\sqrt{\langle( x_i - \langle x_i \rangle)^2
\rangle \langle( x_{i+1} - \langle x_{i+1}\rangle)^2\rangle }}$, takes a
large negative value\cite{comm-braket}.
As in the case of coupled phase oscillators with
different frequencies\cite{Kuramoto},
it is natural to expect that correlations appear when 
the time scale differences $\tau$ become smaller.
Due to chaotic instability, however, desynchronization destroys 
the phase relationship here.  
Switching between low-dimensional correlated motion 
and high-dimensional desynchronized motion, known as
chaotic itinerancy\cite{Ikeda-CI,Kaneko-CI,Tsuda-CI},
appears at various time scales.

Now we will show that the slowest dynamics at $i = L$
can be influenced by the fastest element at $i = 1$.
In order to do so, we have carried out the following numerical experiment:
After the initial transients have died out, at a fixed but otherwise arbitrary
point in the temporal evolution of the system, 
we change the control parameter $r$ of the element $i=1$. 
Then we examine how the dynamics at $i=L$
is influenced by this change 
by measuring a statistical property of $x_L(t)$.

\begin{figure}[htbp]
\begin{center}
\includegraphics[width=8.6cm,height=3.4cm]{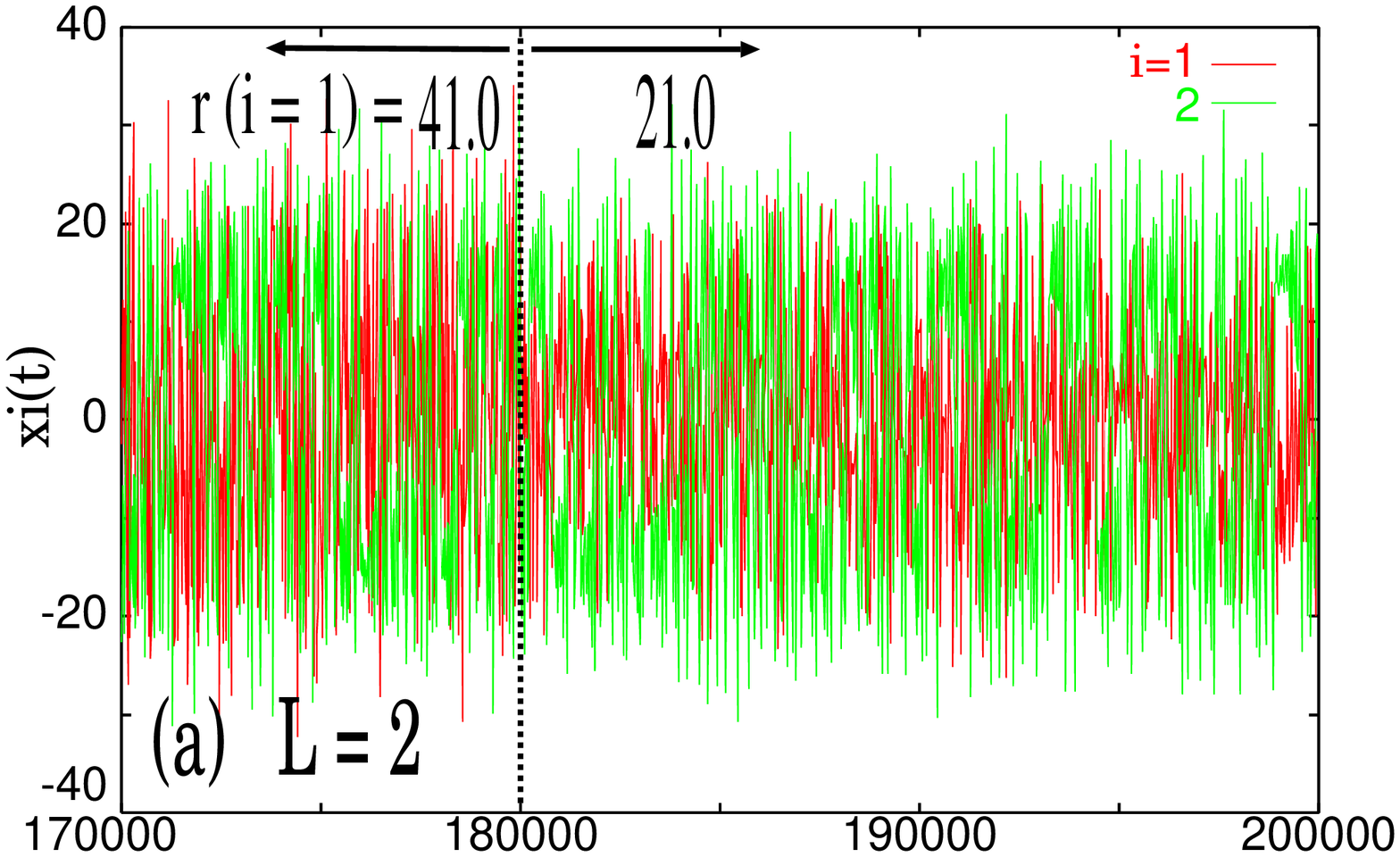} 
\includegraphics[width=8.6cm,height=3.4cm]{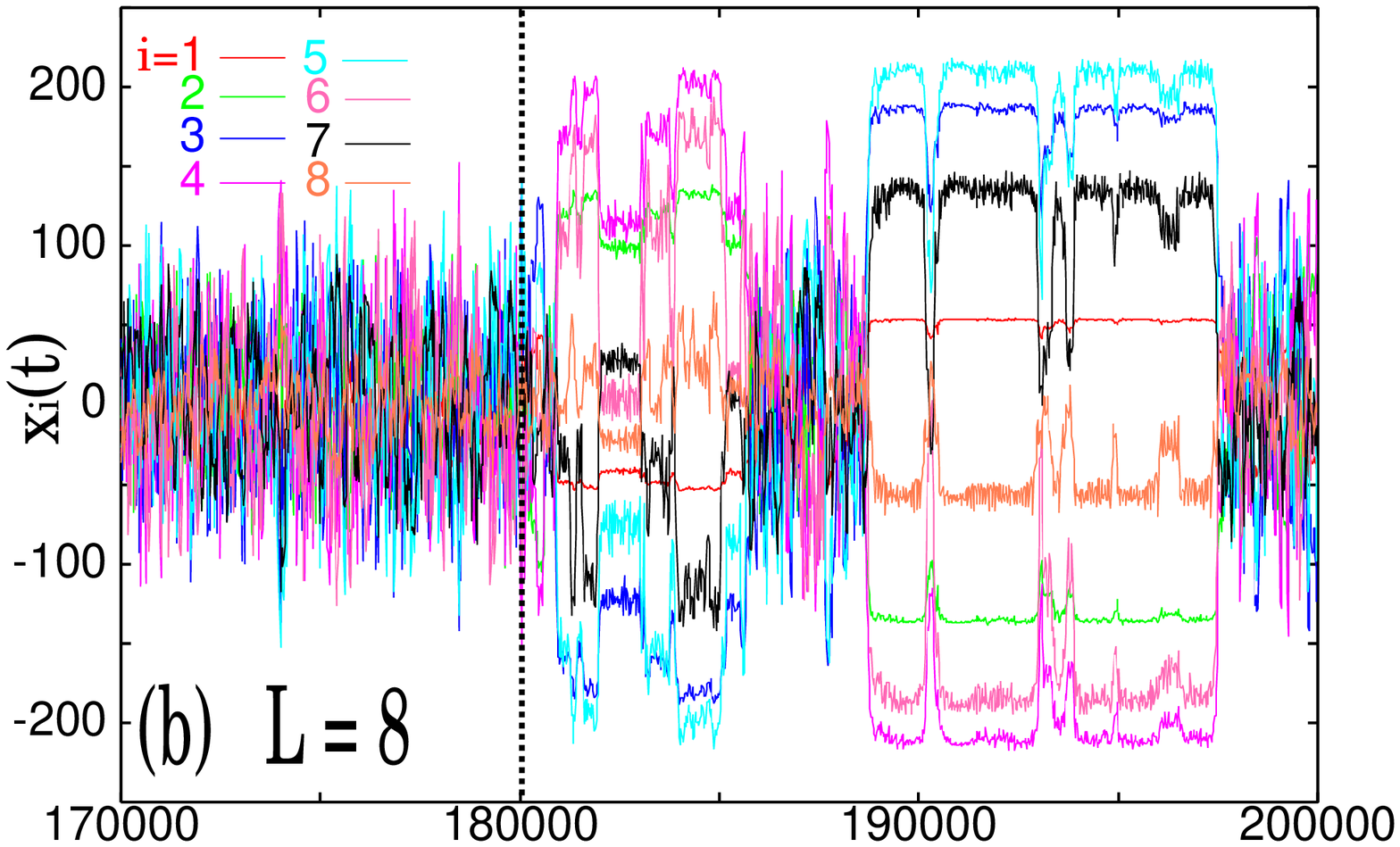} 
\includegraphics[width=8.6cm,height=3.4cm]{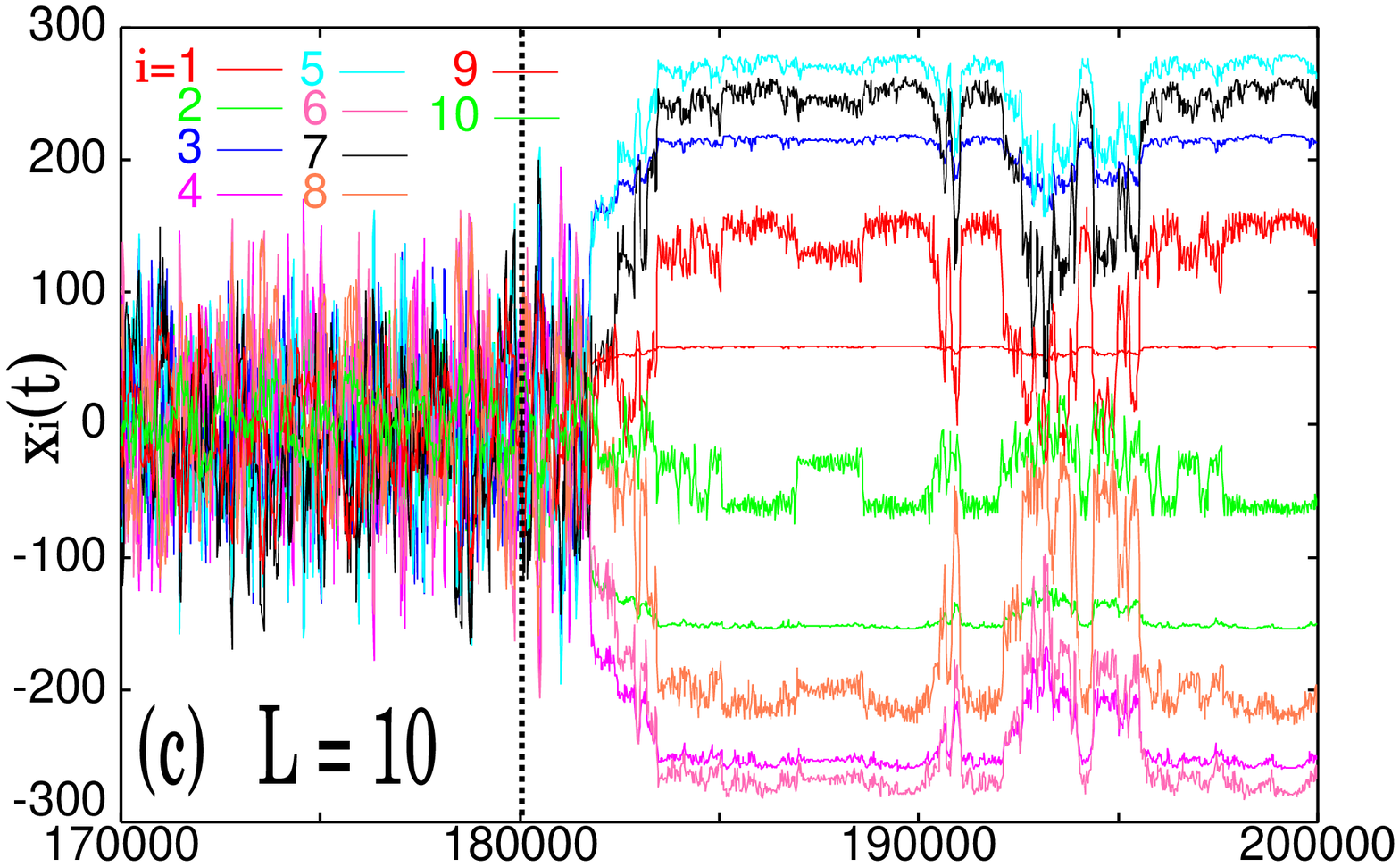} 
\includegraphics[width=8.6cm,height=3.4cm]{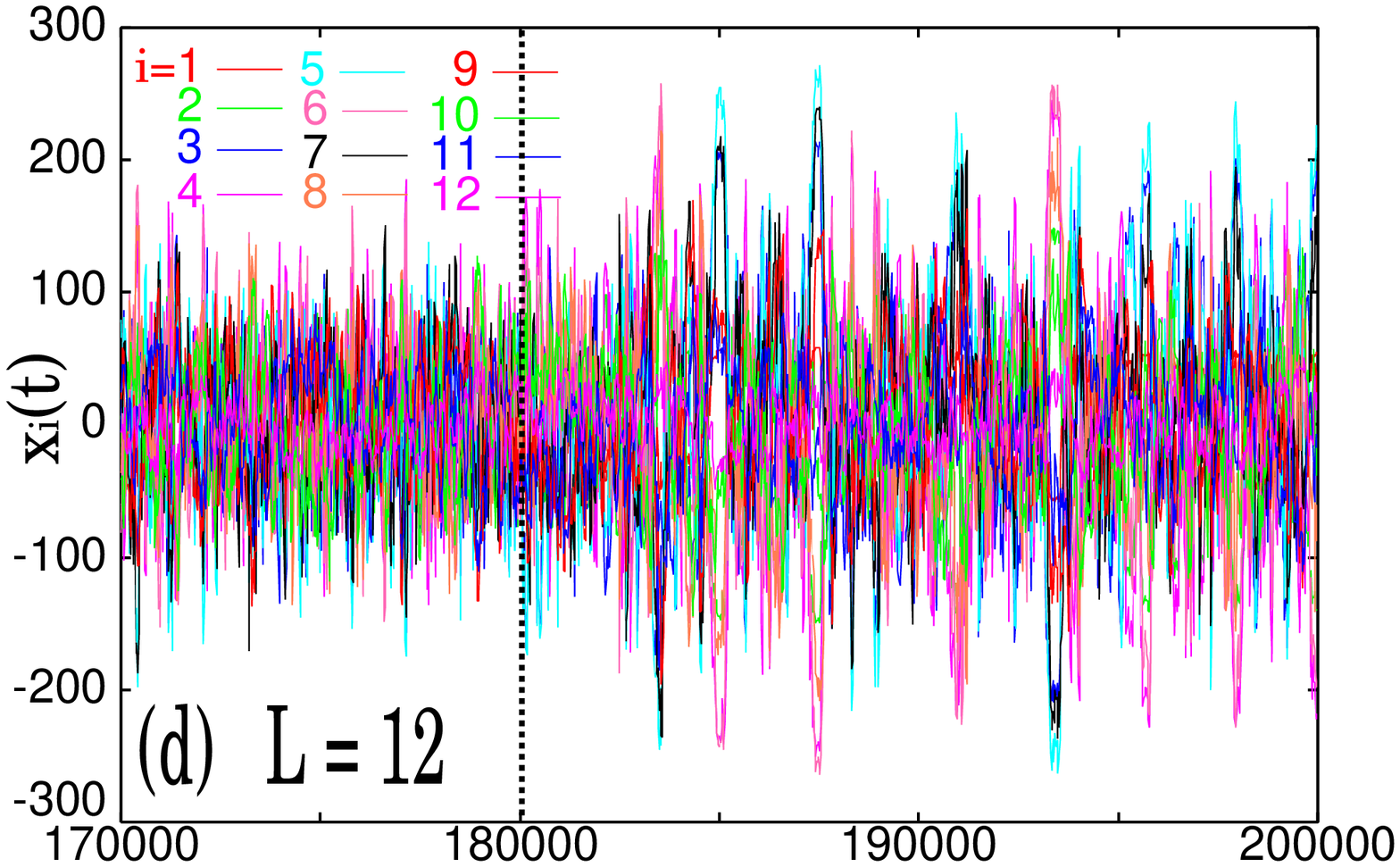} 
\caption{Time series of $x_i(t)$.
At $time = 180000$, the parameter $r$ at $i=1$ is changed from $41$ to
$21$.
$T_{total} = 100$. $(L, \tau) =$
 $(2, 100)$ (a), $(8, 1.93)$  (b),
$(10, 1.67)$ (c), $(12, 1.52)$ (d).
In (b)-(c), novel states with coherent large-amplitude motions appear.
$d = 0.49$. 
}
\label{fig:timeseries-perturb}
\end{center}
\end{figure}

Fig.\ref{fig:timeseries-perturb} shows the time series of $x_i(t)$
for several values of $\tau$
where the parameter $r$ of element $i = 1$ is changed 
from $41$ to $21$ at $time = 180000$.
Here $T_{total} = 100$ and  $(L, \tau) =$
 $(2, 100)$ (a), $(8, 1.93)$  (b),
$(10, 1.67)$ (c), $(12, 1.52)$ (d).
In Fig.\ref{fig:timeseries-perturb}(b)-(c), the change in $r$ leads to  
a novel state with a large amplitude and slow-scale synchronized motion.
The characteristic time scale of this state is about $10^4$.
This is much longer than the time scale of the slowest element $i = L$ which is
about $10^2$ as can be inferred from Fig.\ref{fig:timeseries}(a).
Hence we find that by modifying the fastest dynamics the dynamics of the slowest element
undergoes a  qualitative change.  Such dependence of the slower dynamics on
the control parameter $r$ at $i = 1$ is observed only for 
$6 \leq L \leq 12$ (see Fig.\ref{fig:timeseries-perturb}(a) and
Fig.\ref{fig:timeseries-perturb}(d)).

\begin{figure}[htbp] 
\begin{center}
\includegraphics[width=8.6cm,height=5.5cm]{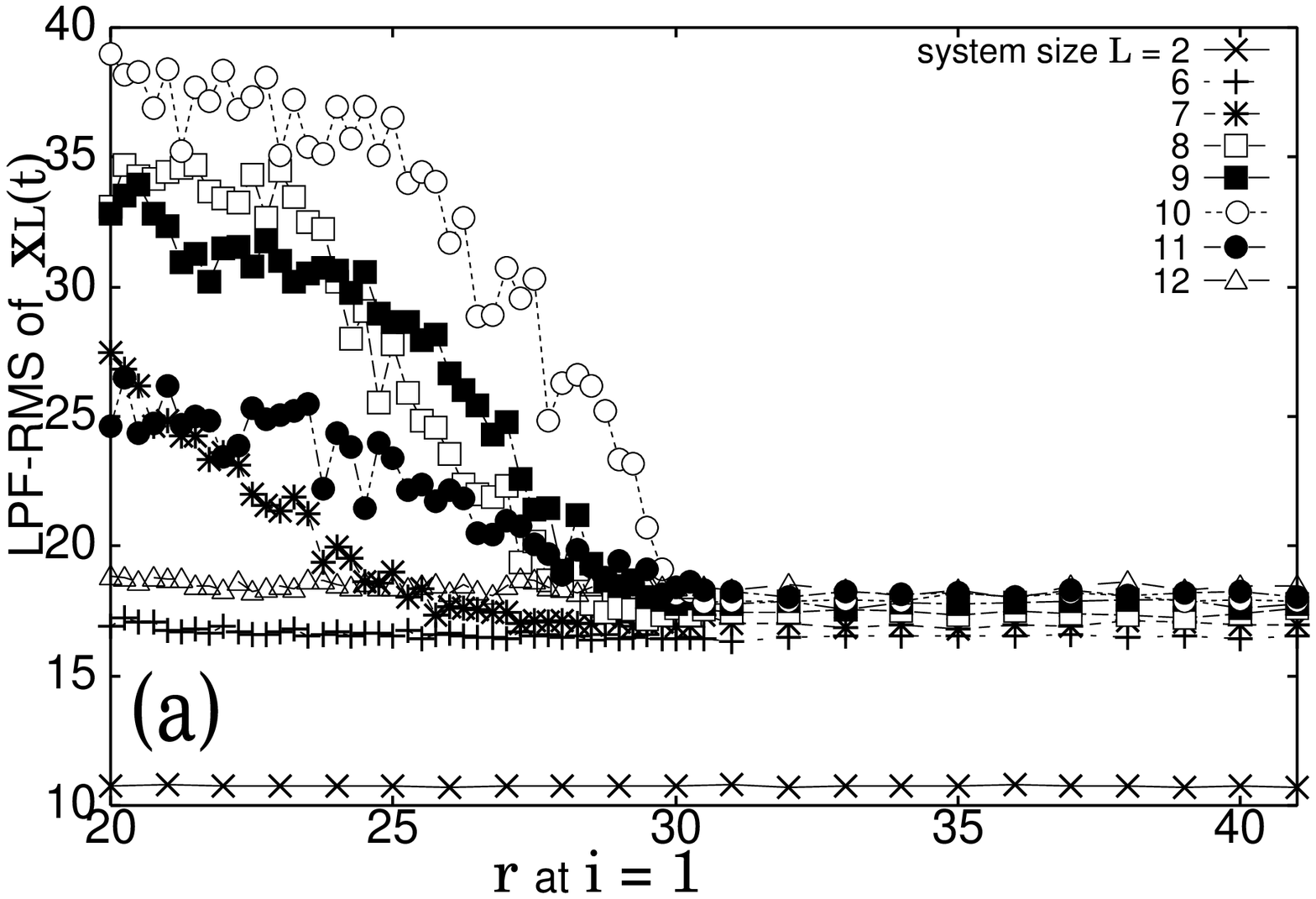} 
\includegraphics[width=7.6cm,height=2.8cm]{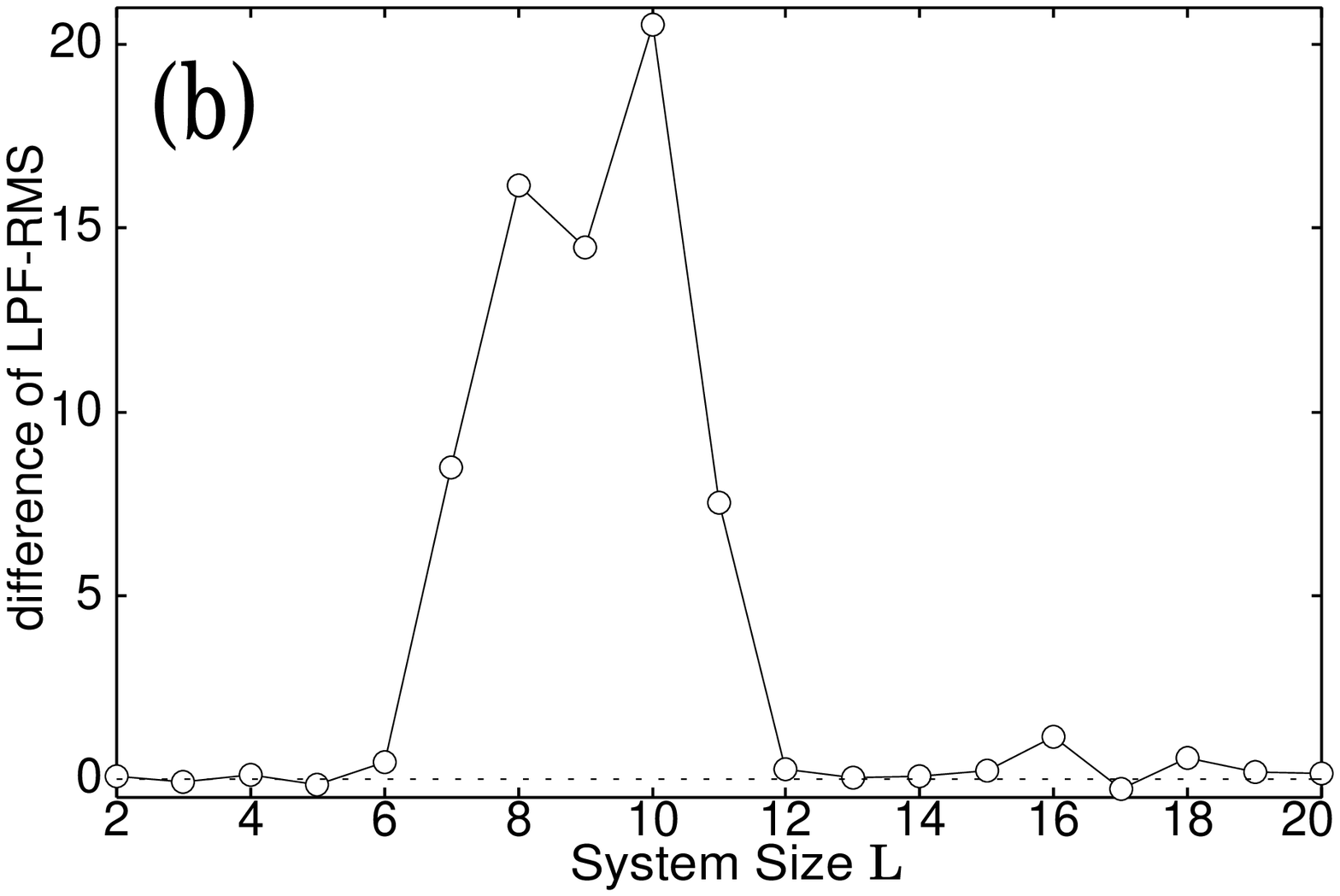} 
\caption{(a) shows the dependence of the
Low-Pass Filtered Root Mean Square (LPF-RMS)
of $x_L(t)$ as a function of
the control parameter $r$ of the fastest element $i = 1$.
$T_{total} = 100$.
The colors indicate different sets of $(\tau, L)$.
Only in the range $7 \leq L \leq 11$
the dependence of slow time scales on fast time scales appears.
(b) shows the system size dependence of the dependence of the slow time scale on the fast time scale.
The ordinate shows the difference of the LPF-RMS of
$x_L(t)$ between the cases of $r = 21$ and $41$
of the element $i = 1$.
}
\label{fig:dep-RMS}
\end{center}
\end{figure}

In order to quantitatively investigate the dependence of the slow time scale on the 
fast time scale,
we low-pass filter $x_L(t)$ by averaging over the faster time scale, i.e.,
$\overline{x_L(t)}\equiv \int_0^{T_a} dt' x_L(t+t')$ and
measure the root mean square of the variation of $\overline{x_L(t)}$ obtaining  the
Low-Pass Filtered Root Mean Square (LPF-RMS)\cite{commemt1}.
In Fig.\ref{fig:dep-RMS}(a) we plot the dependence of the LPF-RMS of  $x_L(t)$
on the control parameter $r$ of the element $i = 1$
for various system sizes $L$.
For sizes $7 \leq L \leq 11$,
the LPF-RMS shows a clear increase as $r$
is decreased from $29$.  To demonstrate the size dependence, we have
plotted the difference of the LPF-RMS of $x_L(t)$ between the cases
$r=41$ and $21$  for  $i = 1$ as a function of $L$.
As clearly can be seen, the dependence of the slowest element on a parameter of the fastest element
only appears for $7 \leq L \leq 11$.
We have also confirmed that this dependence remains
when the system size $L$ is increased, 
(i.e., by increasing $T_{total}$).
Hence the observed dependence is not due to a
finite size effect.
This dependence exists over finite {\it intervals} for the control parameters
$r$ of $i=1$, the coupling strength $d$, and the time scale difference
$\tau$, not only at a 
{\it point} in the parameter space as would be the case for a phase transition
\cite{comm-parameter}.

We also studied a possible
dependence of the fast dynamics on the slow dynamics by
measuring the dependence of the root mean square of the
high-pass filtered value for $x_1(t)$,
$\overline{\overline{x_1(t)}} \equiv x_1(t) - \int_{-T_b/2}^{T_b/2} dt' x_1(t+t')$
on the control parameter $r$ at $i = L$.
No such dependence was found, regardless of the values of
$(\tau, L)$.
Summing up, for values of $\tau$ 
corresponding to $7 \leq L \leq 11$,
the slow dynamics depends on
the fast dynamics, but
the fast dynamics does not depend  on the slow dynamics.
Hence there is an asymmetry in mutual dependence between the 
fast and slow variables.

We now study under what condition
the slower dynamics depends on the faster dynamics. Based on simulations of
the present model for various parameters and also on simulations 
of the coupled R\"{o}ssler equation with different time scales,
we found the following three requirements\cite{KF-KK-Ross}.

First, a strong correlation, either by synchronization or
by an anti-phase relationship, between nearest neighbors in the chain is 
required. 
When there is no such correlation,
the adiabatic approximation for the faster dynamics is valid.
For example, for large values of $\tau$, i.e., for $L \leq 7$ in 
Figs.\ref{fig:dep-RMS}), such coherence is not detectable.

Second, a bifurcation in the fastest dynamics is required
when a control parameter of the fastest element is changed.
In our coupled Lorenz equation with $d = 0.49$, the bifurcation
from a fixed point to chaos occurs 
at $r =r_c \sim 29$ which corresponds to the
bifurcation of a single Lorenz equation at  $r \sim24$.
The bifurcation is required to make possible a switching to a different mode of dynamics.
However, this second condition by itself is not sufficient for transferring the
bifurcated dynamics to
slower elements since the control parameters of these elements are not
shifted.  

For this, a third condition guaranteeing
a cascading transfer of the bifurcation from faster to slower elements is required.
In the examples of Figs.\ref{fig:timeseries-perturb}, 
the control parameter of the fastest element $i=1$ is set to 
$r=21$ allowing for a stable fixed point, while 
the parameters for the other elements $i>1$ are set to  $r=41$ giving chaotic motion.
In this case, as can be seen in Figs.\ref{fig:timeseries-perturb}(b)-(c),
 all elements display chaotic itinerancy between
a highly chaotic state and several ordered states  around fixed points.
As was found in the study of chaotic itinerancy, the switches
from one ordered state to another occur irregularly
through high-dimensional chaotic motion.
In order to distinguish ordered states from chaotic states, 
we measured  the variance over a finite interval,
$v_i(t; t_c(i)) \equiv \frac{1}{t_c(i)}\int_0^{t_c(i)} dt' (x_i(t+t')-x_i(t+T_i+t'))^2$, where
$t_c(i)$ is a constant that is scaled as $t_{c0}\tau ^i$ \cite{comm-Lambda}.  
With the help of this quantity, we can 
roughly roughly estimate whether the element $i$ is in an ordered state 
or not.
If this variance is smaller than its long-term average,
$v_i(\infty) \equiv lim_{T \rightarrow \infty} \frac{1}{T}\int_0^{T} dt' (x_i(t')-x_i(T_i+t'))^2$,
the element is in an ordered state.  From the time series of this
variance we found that the intervals where the chain is in the
ordered state can be rather long 
when there is a dependence of the slower dynamics on the faster dynamics
(i.e., $7 \leq L \leq 11$).

\begin{figure}[htbp]
\begin{center}
\includegraphics[width=6.7cm,height=4.5cm]{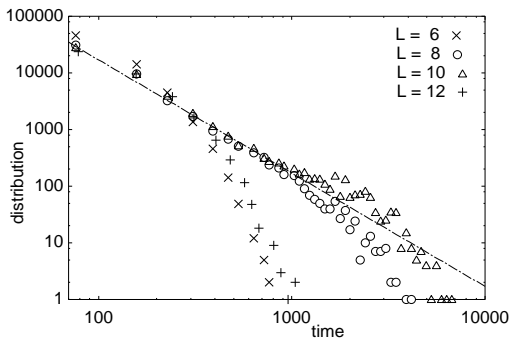} 
\caption{The residence time distribution of ordered states is plotted
for $L = 6, 8, 10, 12$. $T_{total} = 100$.
For $L = 8, 10$, where the propagation from faster to slower elements is
observed, a power-law distribution is observed.
The dotted line with slope $-2$ is plotted for reference.}
\label{fig:cascade}
\end{center}
\end{figure}

In order to check the residence time distribution of the ordered states, 
we have classified states with
$\frac{v_i(t; t_c(i))}{v_i(\infty)} < 0.75$ 
\cite{comm-lamminar} as ordered 
and plotted the results for systems sizes $L = 6, 8, 10, 12$ 
in Fig.\ref{fig:cascade}.
As can be seen, the residence time distribution
follows a power low distribution for $7 \leq L \leq 11$
where the dependence of the slower elements on the fastest element is realized.
This power law distribution allows for the propagation of
bifurcations to ordered states over all elements.  It is reminiscent of
the energy cascade in fluid turbulence 
(as described in the shell model\cite{shell})
and the information cascade in a globally coupled map\cite{Kaneko94}.

When $\tau$ is small,  as is shown in
Fig.\ref{fig:timeseries-perturb}(d),
the chaotic instability 
is large and the sensitivity to the dynamics of neighboring elements
is strongly disturbed by the mixing property. 
Hence the cascade of bifurcations
stops at some element with an intermediate time scale
and cannot be propagated to the slowest element.

In conclusion,
we have shown that a statistical property of a slower element can successively
be influenced by a faster element in a system of coupled chaotic 
elements with different time scales distributed in a power law.  
This propagation of information is realized when there is (i) a 
strong correlation between neighboring elements
such as synchronization or anti-phase oscillation, 
(ii) a bifurcation in the dynamics of the fastest element as its
control parameter is changed, and 
(iii) a cascade propagation of this bifurcation. 
The mutual dependence of the elements in the chain is asymmetric 
in the sense that a change of the fastest element
can influence the slowest element, but that a change of 
the slowest element hardly ever influences the fastest element.
Since the the three requirements were found to be valid for both
the coupled Lorenz equation and the coupled R\"{o}ssler equation, 
we believe that it is reasonable to expect that 
the propagation from faster to slower elements as
described in this letter 
is a universal property of  systems of coupled chaotic dynamics
with distributed time scales.

Biological systems often incorporate dynamics at various time 
scales with
changes at faster time scales sometimes influencing the
dynamics of slower time scales leading to various forms of `memory'.
A cell can e.g. adapt
to an external  condition and maintain its memory over a long time span
through a change of its intra-cellular chemical dynamics.
In a neural system, a fast change in the input is kept as a memory
over a much longer time scale when a
short-term memory is fixated to a long-term memory\cite{brain}.  
In a similar way, a recently proposed dynamical systems theory for evolution proposes
a fixation of
a phenotypic change (by bifurcation) to a slower genetic change\cite{KK-Yomo}.  
The present mechanism for the propagation of a bifurcation from
faster to slower elements will be relevant for the study of such biological systems
and it will be interesting to examine whether the proposed three conditions for 
the dynamics are satisfied as well.
In connection with physics, the possibility of changing the slower dynamics by
controlling a faster element through a cascade process will be important for the
control of turbulence in general.

The authors would like to thank S.Sasa, T.Ikegami, 
T.Shibata, H.Chat\`{e}, M.Sano, I.Tsuda and T.Yanagita for discussions.  
They are also grateful to F. Willeboordse for
critical reading of the manuscript.
This work is partially supported by Grants-in-Aid for Scientific Research
from the Ministry of Education, Science, and Culture of Japan (11CE2006).


\begin{thebibliography}{99}
\bibitem{Haken}
 Haken, {\it Synergetics}  (Springer 1977)
\bibitem{shell}
 M.Yamada and K.Ohkitani, {\it Phys.Rev.Lett.} {\bf 60}, 983 (1988).
\bibitem{Pikovsky}
 M.G.Rosenblum, A.S.Pikovsky and J.Kurths, {\it Phys.} {\it Rev.}  {\it Lett.} {\bf 76}, 1804 (1996). 
\bibitem{comm-braket} 
 $\langle ... \rangle$ denotes the ensemble average.
\bibitem{Kuramoto}
 Y.Kuramoto, {\it Chemical Oscillation, Waves, and Turbulence} (Springer 1984). 
\bibitem{Ikeda-CI}
 K.Ikeda, K.Matsumoto and K.Otsuka, {\it Prog. Theor. Phys. Suppl} {\bf 99}, 295 (1989).
\bibitem{Kaneko-CI}
 K.Kaneko, {\it Physica} {\bf 41D}, 137 (1990).
\bibitem{Tsuda-CI}
 I.Tsuda, {\it Neural Networks} {\bf 5}, 313 (1992).
\bibitem{commemt1} 
 LPF-RMS of  $x_L(t)$ is calculated as
 $\langle(\overline{x_L(t)}-\langle \overline{x_L(t)} \rangle)^2 \rangle$.
\bibitem{comm-parameter}
 For example, for $d = 0.49$, this dependence appears when $r = 29$ at $i = 1$.
\bibitem{KF-KK-Ross}
 K.Fujimoto and K.Kaneko, unpublished.
\bibitem{comm-Lambda}
 In the strange attractor of single Lorenz equation, there are two unstable fixed points.
 When a bifurcation from chaos to fixed point appears, the fixed points are stabilized.
 In the chaotic state, the dynamics of $x_i$ shows transitions from 
 one of the unstable fixed points to the other at time scale $T_i$, accordingly 
 $\frac{1}{t_c}\int_0^{t_c} dt' (x_i(t+t')-x_i(t+T_i+t'))^2$
 takes a large value.
 On the other hand, in the ordered states these fixed points are stabilized,
 accordingly $v_i$ takes a small value.
\bibitem{comm-lamminar}
 We set this condition slightly lower than $\frac{v_i(t; t_c(i))}{v_i(\infty)} = 1$ 
 in order to avoid frequent crossing at $\frac{v_i(t; t_c(i))}{v_i(\infty)} = 1$.
\bibitem{Kaneko94}
 K.Kaneko, {\it Physica} {\bf 77D}, 456 (1994).  
\bibitem{brain}
 For a viewpoint on dynamic memory, see e.g.,
 A.Skarda and W.J.Freeman, {\it Behavioral and Brain Sciences} {\bf 10}, 161 (1987). 
 and \cite{Tsuda-CI}.
\bibitem{KK-Yomo}
 K.Kaneko and T.Yomo, {\it Proc.R.Soc.Rond.} {\bf 257B}, 2367 (2000).
\end{thebibliography}
\end{document}